\documentclass[twocolumn,aps,superscriptaddress,showpacs,nofootinbib,floatfix]{revtex4}
\usepackage{epsfig,bm,feynmf}
\usepackage{graphics}
\usepackage{amsmath}
\usepackage[normalem]{ulem}  
\usepackage[dvips]{color} 

\renewcommand\sout{\bgroup \color{red} \ULdepth=-.5ex \ULset}
\begin{document}


\title{The hadronization time of heavy quark in nuclear matter}


\author{Taesoo Song}\email{song@fias.uni-frankfurt.de}
\affiliation{Institute for Theoretical Physics, Johann Wolfgang Goethe Universit\"{a}t, Frankfurt am Main, Germany}
\affiliation{Frankfurt Institute for Advanced Studies, Johann Wolfgang Goethe Universit\"{a}t, Frankfurt am Main, Germany}

\author{Hamza Berrehrah}\email{berrehrah@fias.uni-frankfurt.de}
\affiliation{Institute for Theoretical Physics, Johann Wolfgang Goethe Universit\"{a}t, Frankfurt am Main, Germany}
\affiliation{Frankfurt Institute for Advanced Studies, Johann Wolfgang Goethe Universit\"{a}t, Frankfurt am Main, Germany}




\begin{abstract}
We study the hadronization time of heavy quark in nuclear matter by using the coalescence model and the spatial diffusion constant of heavy quark from lattice Quantum Chromodynamic calculations, assuming that the main interaction of heavy quark at the critical temperature is hadronization.
It is found that the hadronization time of heavy quark is about 3 fm/c for $2\pi T_c D_s=6$, if a heavy quark is combined with the nearest light antiquark in coordinate space without any correlation between momentum of heavy quark and that of light antiquark which form a heavy meson.
However, the hadronization time reduces to 0.6-1.2 fm/c for charm and 0.4-0.9 fm/c for bottom, depending on heavy meson radius, in the presence of momentum correlation.
Considering the interspace between quarks and antiquarks at the critical temperature, it seems that the hadronization of heavy quark does not happen instantaneously but gradually for a considerable time, if started from the thermal distribution of quarks and antiquarks.
\end{abstract}


\maketitle

\section{introduction}

Relativistic heavy-ion collisions is practically the only way to create extremely hot dense nuclear matter in laboratories.
The Relativistic Heavy Ion Collider (RHIC) and the Large Hadron Collider (LHC), respectively, accelerate heavy nuclei and make collisions up to the energy of 200 GeV and 2.76 TeV.
Such collisions produce strong elliptic flow in semi-central collisions and induce significant energy loss of high $p_T$ particles, which indicate the formation of extremely dense and strongly interacting nuclear matter, so-called strongly-interacting quark-gluon plasma (sQGP).

Searching for the properties of the hot dense nuclear matter is very interesting and also challenging.
Heavy flavor is one of promising probes for the properties.
It has a couple of advantages over other probes.
Firstly, it might have the information about the early stage of nuclear matter, because it is early produced in relativistic heavy-ion collisions.
Secondly, different from light quark, its production is well described in perturbative Quantum Chromodynamics (pQCD)~\cite{Cacciari:2012ny}.

Experimental data from RHIC and LHC show large suppression of nuclear modification factor and strong elliptic flow for heavy flavors~\cite{Adamczyk:2014uip,Tlusty:2012ix,ALICE:2012ab,Abelev:2013lca,Abelev:2014ipa,Adam:2015sza}.
This indicates that heavy flavors also strongly interact with the nuclear matter produced in relativistic heavy-ion collisions.
There have been numerous theoretical studies to explain and describe the experimental data of heavy flavors~\cite{Molnar:2006ci,Zhang:2005ni,Linnyk:2008hp,Uphoff:2011ad,Uphoff:2012gb,Gossiaux:2010yx,Nahrgang:2013saa,Moore:2004tg,He:2011qa,He:2012df,Cao:2011et,Cao:2015eek,Song:2015sfa,Song:2015ykw}.
Most of them take the following steps:
First of all, heavy quark pairs are produced through nucleon-nucleon binary collisions.
Produced heavy quarks and heavy antiquarks then interact with partonic matter in QGP phase.
At the critical temperature for the phase transition, heavy quarks and heavy antiquarks are hadronized into heavy mesons.
Finally, the heavy mesons interact with other hadrons until they freeze out.

The interactions of heavy flavor with nuclear matter have been extensively studied in QGP phase as well as in hadron gas phase.
In the Dynamical Quasi-Particle Model (DQPM), heavy quark interacts with the off-shell partons whose spectral functions are determined from a fit to lattice equation-of-state (EoS)~\cite{Cassing:2009vt}.
It shows that the spatial diffusion constant of heavy quark decreases as temperature approaches the critical temperature~\cite{Berrehrah:2014kba}.
This results are in good agreement with the recent results from lattice Quantum Chromodynamics (lQCD)~\cite{Banerjee:2011ra}.
On the other hand, the spatial diffusion constant of heavy meson in hadron gas has been calculated by using an effective lagrangian, and it decreases with increasing temperature~\cite{Tolos:2013kva}.
Interestingly the diffusion constant of heavy quark in QGP meets that of heavy meson in hadron gas around the critical temperature ($T_c$).
In other words, the diffusion constant is smoothly connected and has the minimum value around $T_c$.
Since the spatial diffusion constant is defined as the squared displacement of a particle per unit time, the small diffusion constant at the critical temperature implies the strong interaction of heavy quark with nuclear matter in phase transition.

Quark coalescence is one of the most popular models to describe the hadronization of partons in nuclear matter~\cite{Greco:2003xt,Greco:2003vf}.
In this model, a pair of quark and antiquark forms a meson, and three quarks and three antiquarks, respectively, form a baryon and an antibaryon.
In this process, heavy quark gains momentum from a coalescence partner or coalescence partners.
Since the spatial diffusion constant is related to the momentum transferred to a heavy quark per unit time, if the hadronization time of heavy quark is given, the diffusion constant can be calculated.

In this study, we calculate the hadronization time of heavy quark in nuclear matter by using the spatial diffusion constant of heavy quark from lQCD calculations and the momentum transfer to heavy quark in the coalescence model.

This paper is organized as follows:
We describe in Sec.~\ref{Ds} the spatial diffusion constant of heavy quark, and in Sec.~\ref{coalescence} the coalescence model.
Combining them, our results are presented in Sec.~\ref{results}, and the summary is given in Sec.~\ref{summary}.

\section{diffusion constant}\label{Ds}

The spatial diffusion constant, $D_s$, is defined as the squared distance per unit time which a particle travels in matter:
\begin{eqnarray}
\langle x_i (t) x_j (t)\rangle=2 D_s t\delta_{ij},
\end{eqnarray}
where $\langle\ldots \rangle$ is ensemble average and the particle is located at ${\bf x}=0$ at $t=0$.
Using the relation in nonrelativistic limit
\begin{eqnarray}
x_i(t)=\int_0^t dt^\prime \frac{p_i(t^\prime)}{M}
\end{eqnarray}
where $M$ is heavy quark mass, we have
\begin{eqnarray}
6D_st=\langle {\bf x}(t)\cdot {\bf x}(t)\rangle=\frac{1}{M^2}\int_0^t dt_1 \int_0^t dt_2 \langle {\bf p} (t_1)\cdot {\bf p} (t_2)\rangle.
\label{dsf}
\end{eqnarray}

On the other hand, the momentum as a function of time is given by random kicks in matter:
\begin{eqnarray}
p_i(t)=\int_{-\infty}^t dt^\prime e^{\eta_D(t^\prime -t)}\xi_i(t^\prime),
\label{momentum}
\end{eqnarray}
where $\eta_D$ is a momentum drag coefficient and $\xi_i$ is the random force which has the correlation
\begin{eqnarray}
\langle \xi_i (t) \xi_j (t^\prime)\rangle=\kappa\delta_{ij}\delta(t-t^\prime),
\end{eqnarray}
where
\begin{eqnarray}
\kappa=\frac{1}{3}\frac{d\langle (\Delta p)^2\rangle}{dt}.
\label{kappa}
\end{eqnarray}

Substituting Eq.~(\ref{momentum}) into Eq.~(\ref{dsf}) and assuming $t\gg \eta_D^{-1}$, we have
\begin{eqnarray}
D_s=\frac{\kappa}{2\eta_D^2 M^2}.
\label{ds1}
\end{eqnarray}

Using the relation,
\begin{eqnarray}
3MT=\langle {\bf p} (t)\cdot {\bf p} (t)\rangle=\frac{3\kappa}{2\eta_D},
\end{eqnarray}
where $T$ is the temperature of matter, the spatial diffusion constant is reexpressed as
\begin{eqnarray}
D_s=\frac{2T^2}{\kappa}.
\label{ds2}
\end{eqnarray}

From Eq.~(\ref{kappa}) and (\ref{ds2}), we finally have
\begin{eqnarray}
\frac{d\langle (\Delta p)^2\rangle}{dt}=\frac{6T^2}{D_s}.
\label{final}
\end{eqnarray}

\section{heavy quark coalescence}\label{coalescence}

The squared transition amplitude for two-particle coalescence is given by
\begin{eqnarray}
|M|^2=|\langle {\bf P}|{\bf p_1p_2}\rangle |^2=\int d^3{\bf x_1} d^3 {\bf x_2} d^3 {\bf x_1^\prime} d^3 {\bf x_2^\prime} \nonumber\\
\times \langle {\bf P}|{\bf x_1x_2}\rangle \langle {\bf x_1x_2}|{\bf p_1p_2}\rangle \langle {\bf p_1p_2}|{\bf x_1^\prime x_2^\prime}\rangle \langle {\bf x_1^\prime x_2^\prime}|{\bf P}\rangle,
\label{transition}
\end{eqnarray}
where two particles with momenta ${\bf p_1}$ and ${\bf p_2}$ form one particle with the momentum ${\bf P}$.
Since instant transition is assumed, there is no time difference between initial and final states in Eq.~(\ref{transition}).

Defining new variables,
\begin{eqnarray}
{\bf R_1}=\frac{{\bf x_1}+{\bf x_1^\prime}}{2},~~~{\bf R_2}=\frac{{\bf x_2}+{\bf x_2^\prime}}{2},\nonumber\\
{\bf r_1}={\bf x_1}-{\bf x_1^\prime},~~~{\bf r_2}={\bf x_2}-{\bf x_2^\prime},~~
\end{eqnarray}

the scalar products in Eq.~(\ref{transition}) are, respectively, expressed as
\begin{eqnarray}
\langle {\bf P}|{\bf x_1x_2}\rangle \langle {\bf x_1^\prime x_2^\prime}|{\bf P}\rangle
=\frac{1}{V}e^{-i{\bf P}\cdot\frac{\bf x_1+x_2}{2}}\psi(\bf x_1-x_2)\nonumber\\
\times e^{i{\bf P}\cdot\frac{\bf x_1^\prime+x_2^\prime}{2}}\psi^*(\bf x_1^\prime-x_2^\prime)~~~~~~~~~~~~\nonumber\\
=\frac{1}{V}e^{-i{\bf P}\cdot\frac{\bf r_1+r_2}{2}}\psi\bigg({\bf R_1}+{\bf R_2}+\frac{{\bf r_1}}{2}+\frac{{\bf r_2}}{2}\bigg)\nonumber\\
\times \psi^*\bigg({\bf R_1}-{\bf R_2}+\frac{{\bf r_1}}{2}-\frac{{\bf r_2}}{2}\bigg),~~~~~
\end{eqnarray}

where $\psi(\bf p)$ is the wavefunction of two particles and $V$ the volume, and

\begin{eqnarray}
\langle {\bf x_1x_2}|{\bf p_1p_2}\rangle \langle {\bf p_1p_2}|{\bf x_1^\prime x_2^\prime}\rangle~~~~~~~~~~~~~~~~~~~~~~~~~~\nonumber\\
=\frac{1}{V^2}e^{i{\bf p_1}\cdot({\bf x_1-x_1^\prime})}e^{i{\bf p_2}\cdot({\bf x_2-x_2^\prime})}=\frac{1}{V^2}e^{i({\bf p_1}\cdot{\bf r_1}+{\bf p_2}\cdot{\bf r_2})}.
\end{eqnarray}

Introducing new variables again,
\begin{eqnarray}
{\bf R}=\frac{{\bf R_1}+{\bf R_2}}{2},~~~{\bf r}=\frac{{\bf r_1}+{\bf r_2}}{2},\nonumber\\
{\bf R^\prime}={\bf R_1}-{\bf R_2},~~~{\bf r^\prime}={\bf r_1}-{\bf r_2},
\end{eqnarray}

the squared transition amplitude is simplified into
\begin{eqnarray}
|M|^2=\frac{(2\pi)^3}{V^2}\delta^3(\bf p_1+p_2-P)\int d^3 {\bf R^\prime} \Phi(\bf R^\prime, k)
\label{transition2}
\end{eqnarray}
where $\bf k=(p_1-p_2)/2$ and $\Phi(\bf R^\prime, k)$ is the Wigner function,
\begin{eqnarray}
\Phi(\bf R^\prime, k)=\int d^3 {\bf r^\prime}e^{i\bf k \cdot r^\prime}\psi\bigg(\bf R^\prime+\frac{{\bf r^\prime}}{2}\bigg)\psi^*\bigg(\bf R^\prime-\frac{{\bf r^\prime}}{2}\bigg).
\label{wigner}
\end{eqnarray}

By using Eq.~(\ref{transition2}), particle yield from coalescence is given by

\begin{eqnarray}
N=V^3\int \frac{d^3{\bf P}}{(2\pi)^3} \frac{d^3{\bf p_1}}{(2\pi)^3} \frac{d^3{\bf p_2}}{(2\pi)^3} f_1 ({\bf p_1}) f_2 ({\bf p_2})|M|^2\nonumber\\
=V\int d^3{\bf R^\prime} \frac{d^3{\bf p_1}}{(2\pi)^3} \frac{d^3{\bf p_2}}{(2\pi)^3} f_1 ({\bf p_1}) f_2 ({\bf p_2})\Phi(\bf R^\prime, k),
\end{eqnarray}

and the differential density by

\begin{eqnarray}
\frac{d(N/V)}{d^3{\bf P}}=\frac{1}{(2\pi)^6}\int d^3{\bf R^\prime} d^3{\bf k}  f_1 ({\bf p_1}) f_2 ({\bf p_2})\Phi(\bf R^\prime, k),
\label{diff}
\end{eqnarray}
where $f_i ({\bf p}_i)$ is the distribution function of particle $i$.
Eq.~(\ref{diff}) clearly shows that the coalescence probability is nothing but the Wigner function which depends on distances between two particles in coordinate and momentum spaces.

Using the wavefunction from the simple harmonic oscillator (SHO),
\begin{eqnarray}
\psi(r)=\bigg(\frac{mk}{\pi^2}\bigg)^{3/8}e^{-\frac{1}{2}\sqrt{mk}~r^2},
\label{waveft}
\end{eqnarray}
where $m$ and $k$ are respectively particle mass and spring constant, and ${\bf r=r_1-r_2}$, we have the Wigner function,
\begin{eqnarray}
\Phi({\bf r, p})=8\exp\bigg[-\frac{r^2}{\sigma^2}-\sigma^2p^2\bigg],
\label{prob}
\end{eqnarray}
where $\sigma=1/\sqrt{mk}$ and ${\bf p=(p_1-p_2)}/2$.

In the case of heavy meson which is composed of partons with asymmetric masses,
the mass $m$ in Eq.~(\ref{waveft}) is substituted by the reduced mass, $\mu=m_1 m_2/(m_1+m_2)$, and $\sigma$ and ${\bf p}$ in Eq.~(\ref{prob}) respectively by $\sigma=1/\sqrt{\mu k}$ and ${\bf p}=(m_2{\bf p_1}-m_1{\bf p_2})/(m_1+m_2)$.

Defining the mean-squared radius of a meson as the average of squared distance of quark and that of antiquark from their center-of-mass~\cite{Song:2013tla},
it is expressed as
\begin{eqnarray}
\langle r_M^2 \rangle=\frac{1}{2}\langle ({\bf R-r_1})^2+({\bf R-r_2})^2\rangle~~~~~~~~~~\nonumber\\
=\frac{1}{2}\frac{m_1^2+m_2^2}{(m_1+m_2)^2}\langle r^2 \rangle=\frac{3}{4}\frac{m_1^2+m_2^2}{(m_1+m_2)^2}~\sigma^2,
\label{radius}
\end{eqnarray}

where ${\bf R}=(m_2{\bf r_1}+m_1{\bf r_2})/(m_1+m_2)$. We note that the coefficient in Eq.~(\ref{radius}) is different from the one in Ref.~\cite{Song:2015sfa} due to the different definitions of ${\bf r}$ and ${\bf p}$.

\section{results}\label{results}

\begin{figure}[b!]
\begin{center}
\includegraphics[width=7.7 cm]{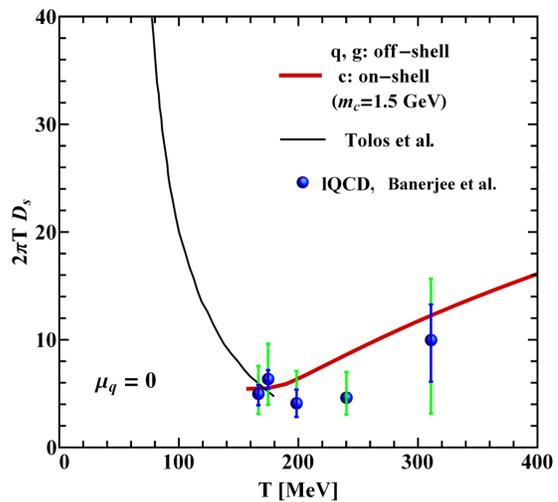}
\caption{(Color online) the spatial diffusion constant
of charm as a function of temperature. The black solid line
below $T=180$ MeV is the hadronic diffusion coefficients~\cite{Tolos:2013kva}, and the red solid line above $T_c
\approx 160$ MeV partonic ones~\cite{Berrehrah:2014kba}. The lattice QCD
calculations are from Ref.~\cite{Banerjee:2011ra}.}
\label{fig:ds}
\end{center}
\end{figure}

Fig.~\ref{fig:ds} shows the spatial diffusion constant of charm as a function of temperature.
Below the critical temperature ($T_c$) the spatial diffusion constants of $D$ meson are calculated by using an effective lagrangian~\cite{Tolos:2013kva}, while those of charm quark are calculated above the critical temperature by using the Dynamical Quasi-Particle Model (DQPM), which reproduce the results from the lattice calculations~\cite{Banerjee:2011ra}.
The figure shows that the diffusion constant of $D$ meson is smoothly connected with that of charm quark around the critical temperature and it has the minimum value there.

Since the spatial diffusion constant is defined as the squared displacement of a particle per unit time, a small diffusion constant implies strong interaction with matter.
In other words, charm strongly interacts with matter near the critical temperature.
It is clearly shown as the strong coupling which increases rapidly near the critical temperature in the DQPM~\cite{Berrehrah:2013mua}.
There is a simple reason for the large strong coupling near the critical temperature:
All partons must be hadronized without exception.

In order to simplify the situation, we prepare a box of which temperature is slightly above $T_c$.
Then the temperature suddenly drops slightly below $T_c$ as happens in relativistic heavy-ion collisions.
In this case most interactions will be hadronization.
As mentioned in the previous section, the momentum transfer to heavy quark is nothing but the momentum of absorbed antiquark in coalescence.
Then the average of squared momentum transfer to heavy quark is expressed as

\begin{eqnarray}
\langle (\Delta k)^2\rangle=\frac{\int d^3{\bf k}~\int d^3{\bf q}~q^2 f_{\bar{q}}(q) f_Q(k)\phi({\bf k,q},r_M)}{\int d^3{\bf k}~\int d^3{\bf q}~ f_{\bar{q}}(q) f_Q(k)\phi({\bf k,q},r_M)},
\label{case1}
\end{eqnarray}
where $f_Q(k)$ and $f_{\bar{q}}(q)$ are respectively the Fermi-Dirac distribution functions of heavy quark and light antiquark at $T_c$ and $\phi({\bf k,p},r_M)$ the momentum part of the coalescence probability in Eq.~(\ref{prob}): $\phi({\bf k,q},r_M)\sim e^{-\sigma^2p^2}$ with ${\bf p}=(m_{\bar{q}}{\bf k}-m_Q{\bf q})/(m_{\bar{q}}+m_Q)$ in the center-of-mass frame of ${\bf k}$ and ${\bf q}$, and $m_{\bar{q}}$ and $m_Q$ being respectively the masses of light antiquark and heavy quark.
In other words, we assume the homogenous distribution of particles in coordinate space.
By using Eq.~(\ref{final}), we can calculate the hadronization time of heavy quark as following:

\begin{eqnarray}
t=\frac{D_s\langle (\Delta k)^2\rangle}{6T_c^2}.
\label{final2}
\end{eqnarray}

\begin{figure}[h]
\centering
\includegraphics[width=0.5\textwidth]{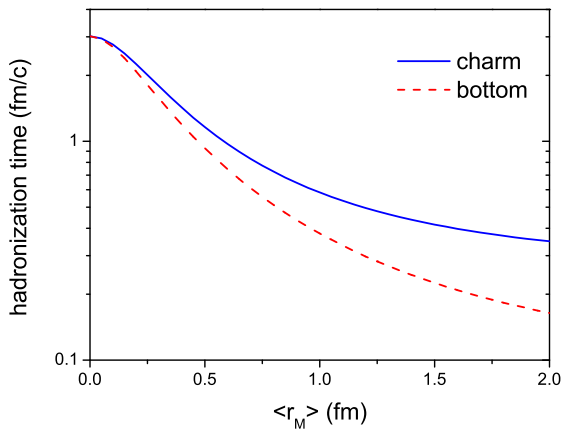}
\caption{(Color online) hadronization times of charm (solid) and bottom (dashed) quarks as functions of heavy meson radius for $2\pi T_c D_s=6$~\cite{Banerjee:2011ra}.}
\label{dt}
\end{figure}

Fig.~\ref{dt} shows the hadronization times of charm and bottom quarks as functions of heavy-meson radius for $2\pi T_c D_s=6$~\cite{Banerjee:2011ra}.
Heavy quark mass $m_Q$ is taken to be 1.5 GeV for charm and 4.5 GeV for bottom, and $m_{\bar{q}}$ and $T_c$ are respectively 0.3 GeV and 160 MeV.

The hadronization time is about 3 fm/c for vanishing radius of heavy meson.
From Eq.~(\ref{prob}) and (\ref{radius}), the vanishing radius implies that a heavy quark does coalescence with the nearest light antiquark in coordinate space.
In this case, there is no correlation between the momentum of heavy quark and that of light antiquark, and the momentum transfer due to coalescence is largest.
Since the momentum transfer per unit time is fixed by $D_s$ from the lattice calculations, hadronization time must be longest from Eq.~(\ref{final2}).

As coalescence radius increases, small relative momentum between heavy quark and light antiquark is favored for coalescence, and the momentum transfer due to the absorption of antiquark becomes small.
It reduces the hadronization time of heavy quark as shown in Fig.~\ref{dt}.
We can also see that the hadronization time of bottom quark is smaller than that of charm quark as the coalescence radius is large.
The large coalescence radius means that only heavy quark and light antiquark which almost comove can be combined in coalescence.
Since the thermal motion of charm quark is larger than that of bottom quark at the critical temperature, the momentum of light antiquark which is combined with charm quark is larger and the momentum transfer to charm quark is also larger.

Assuming that the radius of heavy meson is 0.5-1.0 fm, the hadronization time of charm quark is 0.6-1.2 fm/c and that of bottom quark 0.4-0.9 fm/c.
They are of reasonable time scale and support the results on the spatial diffusion constant from lattice calculations.
Since we neglect the elastic scattering which might give additional momentum transfer to heavy quark, our estimate on the hadronization time of heavy quark is lower limit.

The number density of quark and antiquark at the critical temperature is about 1 ${\rm fm^{-3}}$, and the interspace between them 1 ${\rm fm}$.
Considering the typical size of hadrons, it is highly likely that a heavy quark is combined with the nearest antiquark in coordinate space for hadronization.
In this case, the hadronization time of heavy quark is considerably long and it can be interpreted as following:
Above the critical temperature, quarks and antiquarks have thermal motion, which is random and does not have any correlation between the momentum of quark and that of antiquark.
As energy density decreases, quark and antiquark begin to cluster, and the relative momentum between quark and antiquark becomes small.
In this environment, heavy quark needs shorter time for hadronization as discussed and shown in Fig.~\ref{dt}.
This interpretation suggests that the hadronization of heavy quark does not happen instantaneously, rather requires a considerable time, if started from the thermal distribution of quarks and antiquarks.

\section{summary}\label{summary}

Heavy flavor is one of promising probes for the properties of extremely hot dense nuclear matter created in relativistic heavy-ion collisions.
Since it is massive, heavy flavor is produced mainly through initial nucleon-nucleon binary collisions and exists in the very early stage of relativistic heavy-ion collisions.
After production, heavy quark interacts with partons in QGP phase.
The interactions change the energy-momentum of heavy quark, and it is shown as highly suppressed nuclear modification factor at large transverse momentum and large elliptic flow in semi-central heavy-ion collisions.

From lattice QCD calculations and the Dynamical Quasi-Particle Model, the spatial diffusion constant of heavy quark decreases with decreasing temperature in QGP phase.
On the other hand, the spatial diffusion constant of heavy meson from an effective lagrangian calculations increases with increasing temperature in Hadron gas phase.
Both diffusion constants meet each other around the critical temperature for phase transition and have minimum value there.
Since the spatial diffusion constant is defined as squared displacement per unit time, small diffusion constant near critical temperature implies that strong interactions happen in phase transition.
It is reasonable in respect that without exception all partons should be hadronized in phase transition.
In other words, the small diffusion constant at critical temperature is mostly attributed to hadronization.

The coalescence model has widely been used in describing the hadronization of partons.
In this model, a heavy quark is hadronized into a heavy meson by absorbing a light antiquark nearby in coordinate and momentum spaces.
The absorption transfers the momentum of antiquark to heavy quark.
If the radius of heavy meson is small, heavy quark favors the antiquark near in coordinate space as its coalescence partner.
It allows the coalescence with the antiquark whose momentum is rather far from that of heavy quark.
In this case, the momentum transfer to heavy quark due to hadronization is large.
On the contrary, the momentum transfer is small for the large radius of heavy meson.

Since the spatial diffusion constant is proportional to squared momentum transfer per unit time,
if momentum transfer and diffusion constant are given, the time for the momentum transfer can be calculated.
We have calculated the hadronization time of heavy quark by using the spatial diffusion constant from lattice QCD and DQPM calculations, and the momentum transfer from the coalescence model.
If the radius of heavy meson is extremely small, in other words, a heavy quark does coalescence with any nearest antiquark in coordinate space, the hadronization time is as long as 3 fm/c for $2\pi T_c D_s=6$.
Assuming that the heavy meson radius is 0.5-1.0 fm, the hadronization time is 0.6-1.2 fm/c for charm and 0.4-0.9 fm/c for bottom, where the small (large) radius corresponds to the long (short) hadronization time.
The longer hadronization time for the smaller radius of heavy meson is not so intuitive.
It can be understood from that the small radius in coordinate space allows large momentum transfer to heavy quark for hadronization in the coalescence model.

In principle, there could be additional interactions such as elastic scattering other than hadronization near the critical temperature.
Considering that, our estimate on the hadronization time of heavy quark is a lower limit.

Finally, the consideration of interspace between quark and antiquark at the critical temperature favors the coalescence of heavy quark with the nearest antiquark in coordinate space, and it suggests the gradual progress of heavy quark hadronization for a couple of fermi of time.

\section*{Acknowledgements}
This work was supported by DFG under contract BR 4000/3-1 and
by the LOEWE center "HIC for FAIR". The computational resources have been
provided by the LOEWE-CSC.

\end{document}